\documentclass[usenatbib]{mn2e}

\usepackage{graphicx}
\usepackage{subfigure}
\usepackage{rotating}

\newcommand{\lsim }{{\lower0.8ex\hbox{$\buildrel <\over\sim$}}}
\newcommand{\gsim }{{\lower0.8ex\hbox{$\buildrel >\over\sim$}}}

\usepackage{soul}

\def\Nustar{\emph{NuSTAR}}
\def\Chandra{\emph{Chandra}}

\def\Swiftxrt{\emph{Swift/XRT}}

\def\magnetar{SGR 1745-29}
\def\j174540{CXO J174540.0-290005}

\def\simge{\mathrel{%
  \rlap{\raise 0.511ex \hbox{$>$}}{\lower 0.511ex \hbox{$\sim$}}}}
\def\simle{\mathrel{
  \rlap{\raise 0.511ex \hbox{$<$}}{\lower 0.511ex \hbox{$\sim$}}}}

\newcommand{\Msun}{\ifmmode {M_{\odot}}\else${M_{\odot}}$\fi}
\newcommand{\Lsun}{\ifmmode {L_{\odot}}\else${L_{\odot}}$\fi}
\newcommand{\Rsun}{\ifmmode {R_{\odot}}\else${R_{\odot}}$\fi}

\usepackage{xargs}
\usepackage{color}

\title[The 2013 outburst of a VFXB near Sgr A*]{The 2013 outburst of a transient very faint X-ray binary, 23'' from Sgr A*}
\author[Koch et al.]{E.~W. Koch$^{1,2}$, 
A.~Bahramian$^1$, 
C.~O.~Heinke$^1$, 
K.~Mori$^3$, 
N.~Rea$^{4,5}$, 
N.~Degenaar$^{6,7}$, \newauthor
D.~Haggard$^{8,9}$, 
R.~Wijnands$^4$, 
G.~Ponti$^{10}$, 
J.~M.~Miller$^6$, 
F.~Yusef-Zadeh$^8$, 
F.~Dufour$^{11}$, \newauthor
W.~D.~Cotton$^{12}$, 
F.~K.~Baganoff$^{13}$,
M.~T.~Reynolds$^{6}$\\
$^{1}$ Dept. of Physics, University of Alberta, CCIS 4-183, Edmonton, AB T6G 2E1, Canada; bahramia@ualberta.ca, heinke@ualberta.ca\\
$^{2}$ Unit 5, Physics Program, UBC-Okanagan, 3333 University Way, Kelowna, BC V1V 1V7, Canada; koch.eric.w@gmail.com\\
$^{3}$ Columbia Astrophysics Laboratory, Columbia University, New York, NY 10027 USA\\
$^4$ Astronomical Institute "Anton Pannekoek," University of Amsterdam, Postbus 94249, NL-1090 GE Amsterdam, The Netherlands\\
$^5$ Institute of Space Sciences (CSIC-IEEC), Campus UAB, Torre C5, 2a planta, E-08193 Barcelona, Spain \\
$^6$ Department of Astronomy, University of Michigan, 500 Church Street, Ann Arbor, MI 48109, USA\\
$^7$ Hubble Fellow\\
$^8$ CIERA and Physics \& Astronomy Dept., Northwestern University, 2145 Sheridan Road, Evanston, IL 60208, USA\\
$^9$ CIERA Fellow\\
$^{10}$ School of Physics and Astronomy, University of Southampton, Highfield, Southampton SO17 1BJ, UK\\
$^{11}$ Department of Physics, McGill University, 3600 rue University, Montréal, QC H3A 2T8, Canada\\
$^{12}$ National Radio Astronomy Observatory, 520 Edgemont Road, Charlottesville, VA 22903-2475, USA\\
$^{13}$Kavli Institute for Astrophysics and Space Research, Massachusetts Institute of Technology, Cambridge, MA 02139, USA\\
}

\begin{document}

\date{}

\pagerange{\pageref{firstpage}--\pageref{lastpage}} \pubyear{2014}

\maketitle

\label{firstpage}

\begin{abstract}
We report observations using the Swift/XRT, NuSTAR, and Chandra X-ray telescopes of the transient X-ray source CXOGC J174540.0-290005, during its 2013 outburst.  Due to its location in the field of multiple observing campaigns targeting Sgr A*, this is one of  the best-studied outbursts of a very faint X-ray binary (VFXB; peak $L_X<10^{36}$ erg/s) yet recorded, with detections in 173 ks of X-ray observations over 50 days.  VFXBs are of particular interest, due to their unusually low outburst luminosities and time-averaged mass transfer rates, which are hard to explain within standard accretion physics and binary evolution.  The 2013 outburst of CXOGC J174540.0-290005 peaked at $L_X$(2-10 keV)=$5.0\times10^{35}$ erg/s, and all data above $10^{34}$ ergs/s were well-fit by an absorbed power-law of photon index $\sim1.7$, extending from 2 keV out to $\gsim$70 keV.  We discuss the implications of these observations for the accretion state of CXOGC J174540.0-290005.

\end{abstract}


\section{Introduction}
The majority of low-mass X-ray binaries (LMXBs; neutron stars (NSs) or black holes accreting from low-mass companion stars) are transient systems, showing brief outbursts of X-ray luminosity $10^3-10^9$ times higher than during quiescence.  This transient behaviour is generally understood to be due to thermal-viscous accretion disk instabilities \citep[e.g.][]{Smak84, King98,Lasota01,Coriat12}.  As programs to identify new outbursts from LMXBs in our Galaxy have become more sophisticated, they have identified progressively fainter outbursts.  Monitoring of the Galactic Centre region has uncovered significant numbers of transient outbursts exhibiting peak X-ray luminosities of $10^{34}-10^{35}$ ergs/s, using XMM-Newton \citep{Sakano05}, Chandra \citep{Muno05}, XMM and Chandra \citep{Wijnands06,Degenaar12}, and most recently Swift/XRT \citep{Degenaar09,Degenaar10}. 
A number of LMXB systems remain in this luminosity range persistently or quasi-persistently \citep[e.g.,][]{intZand05,DelSanto07,Campana09,Heinke09b,Degenaar10,ArmasPadilla13b}, so their behaviour is not always transient. 
 Monitoring of X-ray bursts with BeppoSAX's WFC identified a number of thermonuclear X-ray bursts from locations without evidence of persistent emission above $L_X\sim10^{35}$ ergs/s. These ``burst-only'' sources \citep{Cocchi01,Cornelisse02a,Cornelisse02b,Campana09} are also thought to be producing bursts during episodes of low-level accretion.  
X-ray binaries with peak $L_X$ in the range $10^{34}$-$10^{36}$ ergs/s are known as very-faint X-ray binaries (VFXBs). The low time-averaged accretion luminosity of VFXBs is hard to explain in binary evolution models, since it is difficult to reach mass-transfer rates this low via standard LMXB binary evolution within the age of the universe  \citep{King06,Degenaar10,Maccarone13}.

In this paper, we present a detailed study of the 2013 outburst of the VFXB CXOGC J174540.0-290005.  This VFXB was first identified as a transient in a \Chandra\ observation in 2003 at $L_X=3.4\times10^{34}$ ergs/s, 23'' from Sgr A*, while the sum of other deep \Chandra\ observations indicated a quiescent (2-8 keV) $L_X$ upper limit of $4\times10^{31}$ ergs/s \citep{Muno05}. A Swift-XRT monitoring campaign observed a 2-week-long outburst from a nearby transient starting on 20 Oct. 2006 \citep{Kennea06,Kennea06b}, reaching a peak $L_X$ of $2.3\times10^{35}$ ergs/s on 22 Oct. 2006 \citep{Degenaar09}.  A Chandra observation on Oct. 29, 2006 identified this transient as \j174540\  \citep{Degenaar09}.  
\citet{Wang06} used the PANIC near-infrared camera to observe CXOGC J174540.0-290005 on 30-31 October 2006, finding no evidence of an infrared counterpart within the 90\% confidence area given by \citep{Muno05}. However, as \citet{Degenaar09} note, these observations were taken after \j174540\ had dropped below Swift's detection limit.

The identification of outbursts reaching above $10^{36}$ ergs/s from VFXBs that have also shown fainter outbursts indicates that low-luminosity outbursts are a behaviour, not necessarily indicating a fundamentally different class of objects \citep[e.g.][]{Degenaar10}.  Furthermore, VFXB outbursts can be produced by a range of objects, including high-mass X-ray binaries \citep{Torii98}, symbiotic X-ray binaries \citep{Masetti07}, and cataclysmic variables \citep{Mukai08,Stacey11}.  However, high-mass X-ray binaries or M-giants can be ruled out for most X-ray sources in the Galactic Centre \citep{Laycock05,Mauerhan09,DeWitt10}, cataclysmic variables cannot reach the peak  luminosities attained by most VFXBs, and neither produce X-ray bursts as have now been seen from numerous VFXBs \citep[][]{Cornelisse02a}, so LMXBs appear to be the primary source type seen.  In some cases VFXB peak luminosities may be due only to geometric effects (observing the system edge-on, so that most of the true luminosity is hidden by the accretion disk, e.g. \citealt{Muno05b}), though this is unlikely to account for the majority of the systems \citep{Wijnands06}.

Some VFXBs have high-quality X-ray spectra, and in some cases display spectral evolution, during an outburst. 
\citet{ArmasPadilla13b} identify clear soft thermal components in XMM-Newton spectra of two persistent VFXBs at $L_X\sim3\times10^{34}$ ergs/s, and a relatively soft power-law (photon index $\Gamma=2.5$) for one at $L_X\sim10^{35}$ ergs/s.  \citet{ArmasPadilla11} found a softening of the 0.5-10 keV X-rays for the transient VFXB  XTE J1719-291, over the $L_x$ range from $7\times10^{35}$ ergs/s down to $10^{34}$ ergs/s, from one XMM-Newton and several Swift/XRT and RXTE observations.  Swift J1357.2-0933, believed to be a black hole \citep{CorralSantana13}, showed spectral softening in Swift/XRT data as it declined \citep{ArmasPadilla13a}, and its XMM-Newton spectrum revealed evidence for a soft component \citep{ArmasPadilla13d}.   \citet{Degenaar12} find significant softening in the persistent VFXB XMMU J174554.1-291542 in multiple XMM observations covering $L_X\sim$(4-10)$\times10^{33}$(d/6 kpc)$^2$ ergs/s, although they suggest (from spectral hardness and an infrared counterpart) that this may be a symbiotic star. A softening during the last stages of outburst decay is commonly observed among luminous X-ray binary outbursts, both from black holes (both stellar-mass black holes, \citealt{Wu08, Plotkin13}, and also in AGN at similar Eddington fraction, \citealt{Constantin09, Gultekin12}), and in neutron stars, where emission from the NS surface may be relevant \citep{Jonker03,Degenaar13a,Linares13}.  For instance, \citet{Bahramian14} observe hardening during the outburst rise of a bright NS transient, and connect the spectral evolution to the changing optical depth of a hot Comptonizing atmosphere \citep{Deufel01}.

In anticipation of possible interactions between the Sgr A* supermassive black hole and the infalling gas cloud G2 \citep{Gillessen12}, X-ray monitoring campaigns of the Galactic Center have been undertaken using Swift-XRT, \Chandra, and NuSTAR, with which we have serendipitously followed a full outburst of CXOGC J174540.0-290005.   The outburst was detected by NuSTAR \citep{Dufour13}, confirmed and tracked by SWIFT/XRT \citep{Degenaar13e}, and confidently associated with \j174540\ using one of several Chandra observations \citep{Heinke13_atel}.  We have simultaneous observations with the Karl G. Jansky Very Large Array (VLA), which we use to place a radio upper limit on \j174540.  The frequency of our X-ray observations during the outburst, along with their sensitivity and wide energy range, allows us to study a very faint LMXB outburst in more detail than is usually possible.  In \S 2 below, we describe the data and spectral analysis from these instruments.  In \S 3, we provide an accurate position for the transient, discuss its lightcurve, spectrum, and spectral variations, and compare its behaviour to that in previous outbursts.  In \S 4 we discuss how our results relate to analyses of low-luminosity NS X-ray binary spectra.  Unless otherwise specified, all errorbars are at 90\% confidence for one parameter of interest.

\section{Analysis of Individual Instruments} 
\label{sec:data_reduction}

This section describes the data and analysis for each instrument, culminating with spectral fits summarized in Table  \ref{tab:group_fits} for the X-ray data.

\begin{figure*}
\includegraphics[scale=0.35]{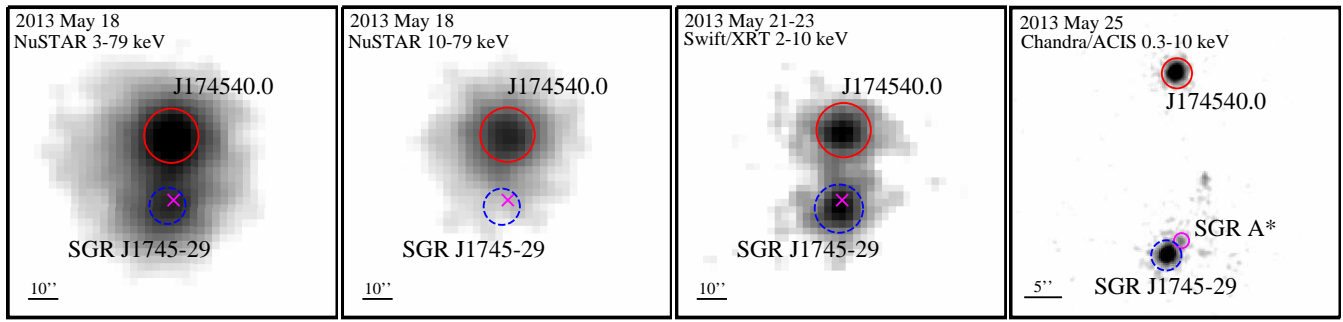}
\caption{Images of CXO J174540.0-290005 (red circle) from different telescopes (NuSTAR, Swift/XRT, and Chandra/ACIS). The blue dashed circle indicates SGR 1745-29 (clearly visible in all images except NuSTAR 10-79 keV), and the magenta circle or cross indicates the position of Sgr A* (visible only in the Chandra image).}
\label{fig:images}
\end{figure*}

\subsection{NuSTAR} 
\label{sub:nustar}
Since the discovery of the new magnetar \magnetar\ near the Galactic Center \citep{Kennea13, Mori13}, \Nustar\ has performed 13 target-of-opportunity observations between 4/26/2013 and 8/13/2013.  \Nustar\ detected \j174540\ twice, on 5/18/2013 (ObsID 2013008) and 5/27/2013 (ObsID 2013010) with exposure times of 39.0 and 37.4 ksec. 
In this section, we present \Nustar\ spectral analysis of \j174540\ using the two detections, and upper limits derived before and after, on 5/11/2013 (ObsID 2013006) and 6/14/2013 (ObsID 2013012).

\Nustar\ consists of the two co-aligned X-ray telescopes (FPMA and FPMB) with an energy band of 3-79 keV and spectral resolution of 400 eV (FMHM) at 10 keV \citep{Harrison13}. Although \Nustar\ was only able to partially resolve \j174540\ and \magnetar\ with its 18" (FWHM) angular resolution, we carefully removed the contamination from \magnetar\ in our spectral analysis. \Nustar\ data processing and analysis were performed with the \emph{NuSTAR Data Analysis Software (NuSTARDAS)} v.1.2.0.\footnote{http://heasarc.gsfc.nasa.gov/docs/nustar/analysis/} We analyzed \Nustar\ spectra from FPMA and FPMB separately, but later we jointly fit them using XSPEC.\footnote{http://heasarc.gsfc.nasa.gov/docs/xanadu/xspec/} 

Prior to our spectral analysis, we applied an astrometric correction to \Nustar\ event files by registering \magnetar\ to its Chandra position RA = 17h45m40s.169 and DEC =  -29deg00'29".84 (J2000) \citep{Rea13} in the 3-10 keV band. For the two \Nustar\ observations, we extracted source photons from a 15" radius circle around the Chandra position of \j174540\ at RA = 17$^h$45$^m$40$^s$.07  and DEC = -29$^{\circ}$00$\farcm$05$\farcs$.8 (J2000).  (See Figure \ref{fig:images}.) The source extraction circle was chosen to obtain high signal-to-noise ratio source spectra by minimizing the contamination from  \magnetar, located $\sim20$" away from \j174540, although the contamination was still substantial. We therefore extracted background spectra from pre- and post-outburst \Nustar\ observations on 5/11/2013 (ObsID: 2013006) and 6/14/2013 (ObsID: 2013012) using the same source extraction region, after verifying that we did not detect \j174540\ in these data.  The pre- and post-outburst spectra are identical within statistical uncertainties, so we took the average of the two spectra and used it as a background spectrum. Since the X-ray flux variation of \magnetar\ was small during the \j174540\ outburst \citep{Kaspi14}, the contamination due to  \magnetar, as well as diffuse background in the Galactic Center, should be  subtracted in our analysis. We binned the \Nustar\ spectra using {\it grppha} to attain a minimum of 30 counts per bin. After background subtraction, the net \Nustar\ 3-79 keV count rates are $0.103\pm0.002$ and $0.108\pm0.002$ cts/sec for FPMA and FPMB respectively. 

To measure upper limits on \j174540's flux for ObsIDs 2013006 and 2013012, we performed a similar analysis, taking background spectra from the observations before and after the outburst, respectively.  For these observations, the flux normalization was consistent with zero at the 1-$\sigma$ level. 

\begin{figure*}
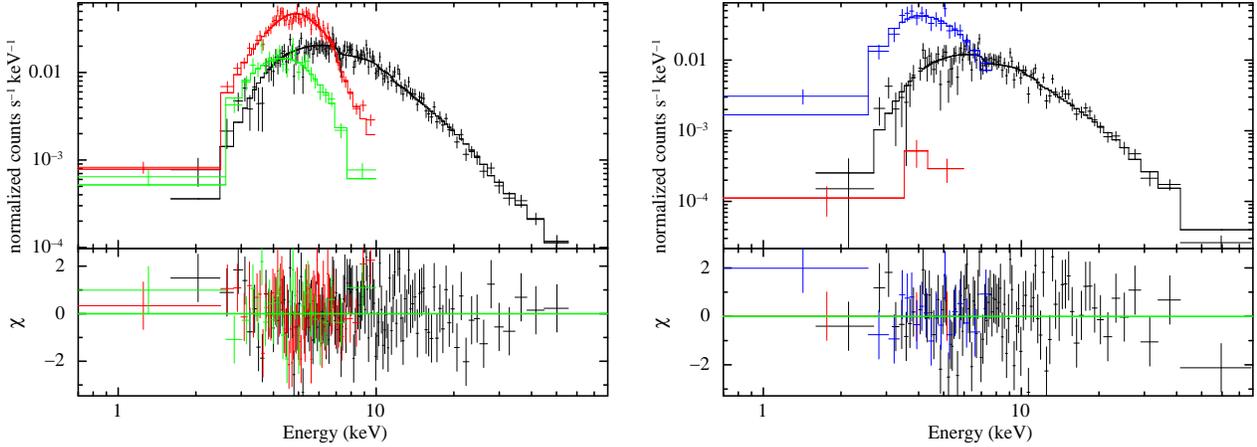

\includegraphics[angle=270, scale=0.33]{xspec_panel1.eps}
\includegraphics[angle=270, scale=0.33]{xspec_panel2.eps}
\caption{Key spectra of \j174540\ fit with an absorbed powerlaw (see text for details), showing data and model (top) and residuals in units of $\Delta \chi^2$ (bottom). {\it Left:} NuSTAR ObsID 2013008 (black), Chandra/ACIS zeroth-order spectrum for ObsID 15040 (red), Chandra/ACIS ObsID 14703 (green).  {\it Right:} NuSTAR ObsID 2013010 (black), Swift/XRT spectra from merged ObsIDs 91736040-42 (blue), and Chandra/ACIS ObsID 14946 (red).
}
\label{fig:xspec}
\end{figure*}

We performed spectral analysis of the \Nustar\ detections in 2-79 keV using HEASOFT 6.13 and XSPEC 12.8.1 with CALDB 20130509, and considered an absorbed power-law model to fit the data, using the TBABS absorption model with \citet{Wilms00} abundances and \citet{Verner96} cross-sections.  Both \Nustar\ modules (A \& B) were considered as a single data group and were jointly fit. 
We find that the \Nustar\ spectra are well fit (reduced $\chi^2$ of 1.01 and 0.99 for 2013008 and 2013010, respectively; see Table 2, Figure \ref{fig:xspec}) with absorbed power-law models, with photon indices of 1.73/1.72 in each, though the flux decreased by a factor of two in nine days.  These measurements are of particular interest as they are the first \Nustar\ spectra from an LMXB with an $L_X$ of a few $\times10^{35}$ ergs/s.  The \Nustar\ spectra do not show obvious evidence (e.g. strong residuals to the power-law fit) of curvature (apart from that attributable to $N_H\sim2.6\times10^{23}$ cm$^{-2}$ at the soft end); e.g. Fig. 2.  We note that the inferred unabsorbed 2-79 keV flux is 3.1 times the unabsorbed 2-10 keV flux, which agrees with \citet{intZand07}'s estimate that $L_{\rm bol}\sim$3$L_X$(2-10 keV).

Fitting the NuSTAR spectra with a physically motivated Comptonization model ({\it TBABS*COMPTT} in XSPEC, \citealt{Titarchuk94}) also gives a reasonable fit (reduced $\chi^2$ of 1.01 and 0.99 respectively; see Table 3).  The brighter \Nustar\ spectrum, from ObsID 2013008 on May 18, gives only a lower limit (95\% single-sided confidence interval) of 40 keV for the Comptonizing electron temperature, at $L_X$(2-10 keV)=$3.5\times10^{35}$ ergs/s.  The fainter \Nustar\ spectrum, at only $1.8\times10^{35}$ ergs/s, however, shows weak evidence of a spectral turnover, measuring $kT_e$=$14^{+96}_{-3}$ keV (90\% confidence range) for the Comptonizing electron temperature.  Although this fit suggests a relatively low Comptonizing electron temperature, it is not highly constraining--an electron temperature of 100 keV is still allowed.

The \Nustar\ data do not show strong evidence of lines from \j174540.  We tested a fit including an iron line with $\sigma$ fixed to 0.1 keV, and energy allowed between 6 and 7 keV.  For ObsID 2013008, we found that the best-fit line energy at 6.8 keV gave an improvement of only 3.4, losing 2 degrees of freedom, in the $\chi^2$ of 472.7 for a power-law fit.  An F-test indicates that there is a 20\% likelihood of obtaining such an improvement by chance, and therefore that there is not strong evidence for an iron line in this spectrum.  ObsID 2013010 gives similar results.

\begin{figure*}
\includegraphics[scale=0.8]{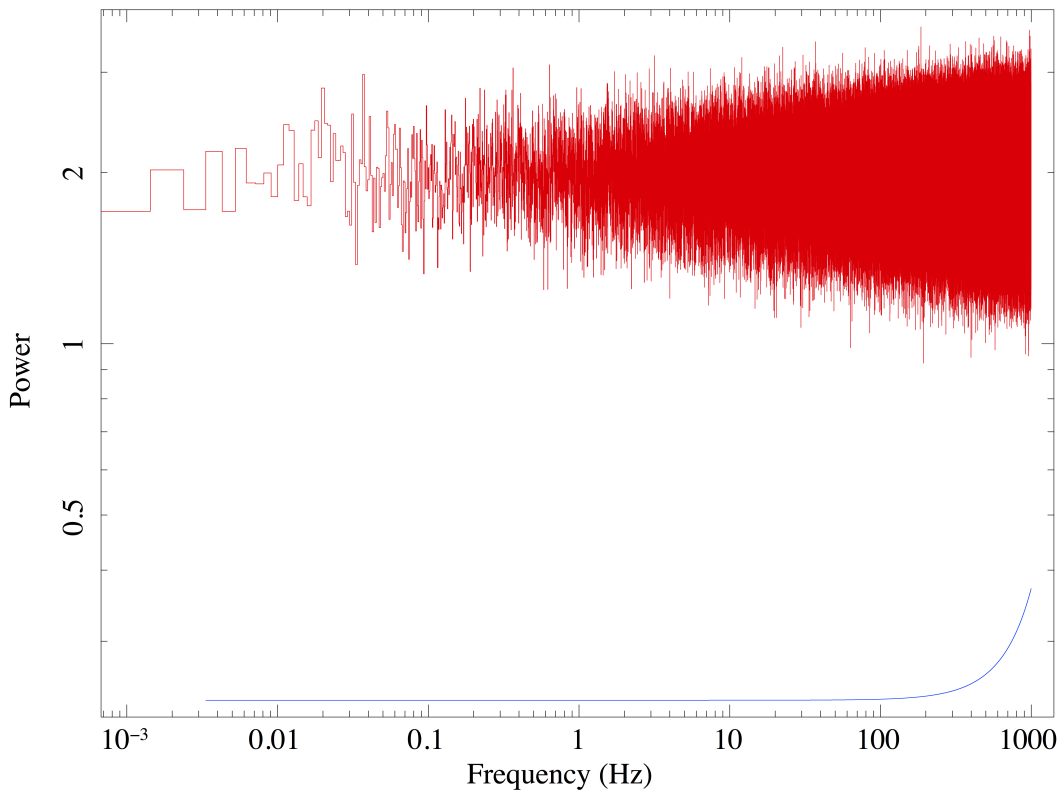}
\caption{Power spectrum (red) of \j174540\ from \Nustar\ ObsID 2013008, showing no evidence for pulsations.  Blue solid line indicates the pulsed fraction limit as a function of frequency, in units of 100\%(0.1=10\%), with the pulsed fraction upper limit gradually rising above 100 Hz.  }
\label{dps}
\end{figure*}

Some models (e.g. \citealt{Cumming01}) suggest that high mass transfer rates 
may screen the magnetic field of accreting NSs in LMXBs, motivating searches for pulsations in 
low-mass-transfer-rate LMXBs, such as VFXBs.
We searched for periodic and quasi-periodic signals in the NUSTAR data
(sensitive to signals between the Nyquist frequency  and half of
the frequency resolution of our \Nustar\ data; \citealt{vanderKlis89}). We
studied the source power spectra performing Fast Fourier Transforms
(FFTs; see Fig.\,\ref{dps}) using the {\tt Xronos} analysis software.
 We did not find any periodic or quasi-periodic signal in any of
the \Nustar\ observations, nor when considering smaller chunks of data (we
 accounted for the number of bins searched, and the different
degrees of freedom of the noise power distribution in the non-detection level;
see \citealt{Vaughan94}; \citealt{Israel96}) .

We computed the 3$\sigma$ upper limits on the sinusoidal semi-amplitude
pulsed fraction ($PF$), computed according to \citet{Vaughan94}
and \citet{Israel96}. The deepest limits were obtained using
observation ObsID 2013008 during the LMXB outburst. The pulsed fraction limit in
the 3--79\,keV energy band is $<$22\% for periods below 100 Hz, rising to $\simle$30\% for 800 Hz 
(see Fig.\,\ref{dps}).  The derived pulsed fraction limits do not consider non-detections due to
Doppler smearing, which may be significant in an LMXB such as this, 
but we do not have sufficient signal to perform a full acceleration search.
The typical pulse fractional amplitudes for accreting millisecond X-ray pulsars are a few percent, 
although some reach up to 30\% \citep{Patruno12}.  Thus, our upper limits 
do not strongly constrain whether \j174540\ is an accreting millisecond pulsar.

\subsection{Swift/XRT} 
\label{sub:swift_xrt}

The region surrounding Sgr A* was observed almost daily using the Swift X-ray Telescope (Swift/XRT, effective energy range 0.5-10 keV, \citealt{Burrows05}) from May to July 2013 as part of the Sgr A* Swift Monitoring Program \citep{Degenaar13c}. This program enabled the detection of the new magnetar SGR J1745-29 \citep{Mori13,Kennea13}, and of \j174540\ in outburst 20'' north of Sgr A*. The outburst of \j174540\ was first detected by \textit{Swift} on 15 May, and was visible to Swift for $\sim$3 weeks. We analyzed 38 \textit{Swift/XRT} observations in photon counting (PC) mode from 11 May to 13 July 2013, which represents about 38 ks of observation time. Observations were summed into several groups, based on similar count rates, to improve constraints during spectral fitting (see Table \ref{tab:observs}). We also used archival data from 15 to 31 October 2006 when this object was last observed in outburst \citep{Degenaar09}, for comparison. This represents an additional 15 data sets and about 23 ks of observation time.

The data was reduced and analyzed using HEASOFT 6.13 and FTOOLS\footnote{http://heasarc.gsfc.nasa.gov/ftools/} \citep{Blackburn95} following the \textit{Swift/XRT} analysis threads\footnote{http://www.swift.ac.uk/analysis/xrt/}. The FTOOLS routine \textit{xselect} was used to extract a spectrum from a circular source with a 10'' radius and background region, near the source, from each group of observations. Due to the close proximity of \magnetar\ in outburst, we reduced the extraction radius (down to 6'') as the luminosity decayed, to reduce the contaminating counts from \magnetar. 
We did not correct for pile-up, since the 2013 outburst peak count rate only reached 0.19 count s$^{-1}$, well below the lower limit of 0.5 count s$^{-1}$ where pile-up becomes significant \footnote{http://www.swift.ac.uk/analysis/xrt/pileup.php}.  Similarly, the 2006 outburst reached a peak count rate of 0.09 count s$^{-1}$, and did not require pileup corrections. The FTOOLS routine \textit{xrtmkarf} was used to create ancillary response function (ARF) files for each group. Spectral analysis was performed using XSPEC 12.8.0 \citep{Arnaud96}.

The count rates, or count rate limits, for each Swift/XRT observation (see Table 1) were derived by counting the photons within a specified radius, 14'' for the high-flux measurements, or 6'' for the lowest-flux measurements.  We measured backgrounds from a region at roughly the same distance from \magnetar, except for the lowest-flux measurements where an annulus around \magnetar\ at the same distance as the (excluded) source extraction region was used.
From the Poisson upper bounds at 95\% confidence \citep{Gehrels86}, we calculate upper bounds on the count rates 
 within a 10'' radius of the object's position, using an enclosed energy of 50\% within this region \citep{Moretti05}. 
We used \textit{PIMMS}\footnote{http://asc.harvard.edu/toolkit/pimms.jsp} with the corrected count rate and the model parameters found during spectral fitting (see \S\ref{sub:outburst} and Table \ref{tab:group_fits}) to calculate an unabsorbed flux in the 2-10 keV energy range. 
The luminosities, and luminosity upper limits, assumed a distance of 8.0 kpc to \j174540.

\begin{figure*}
\includegraphics[scale=0.41]{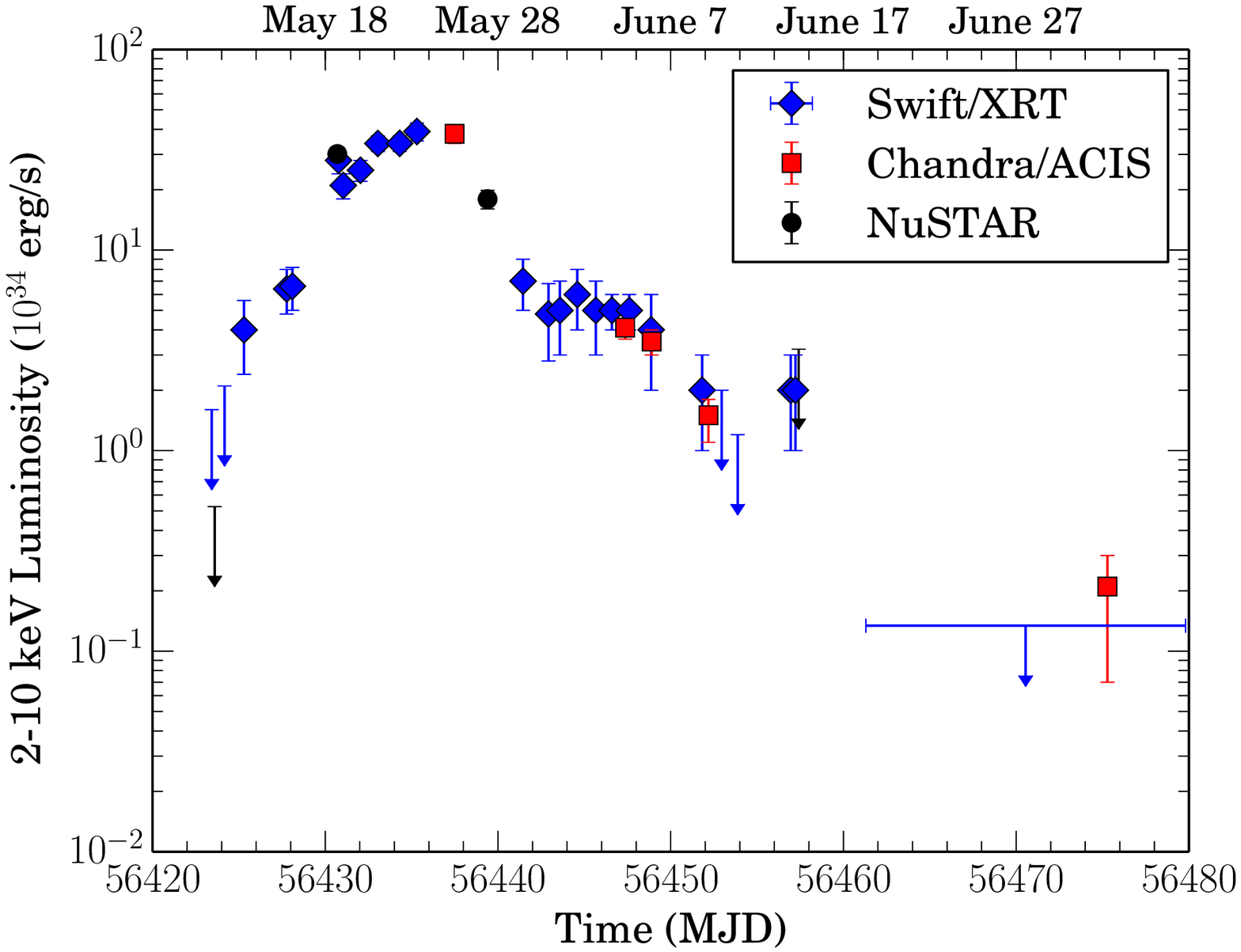}
\includegraphics[scale=0.41]{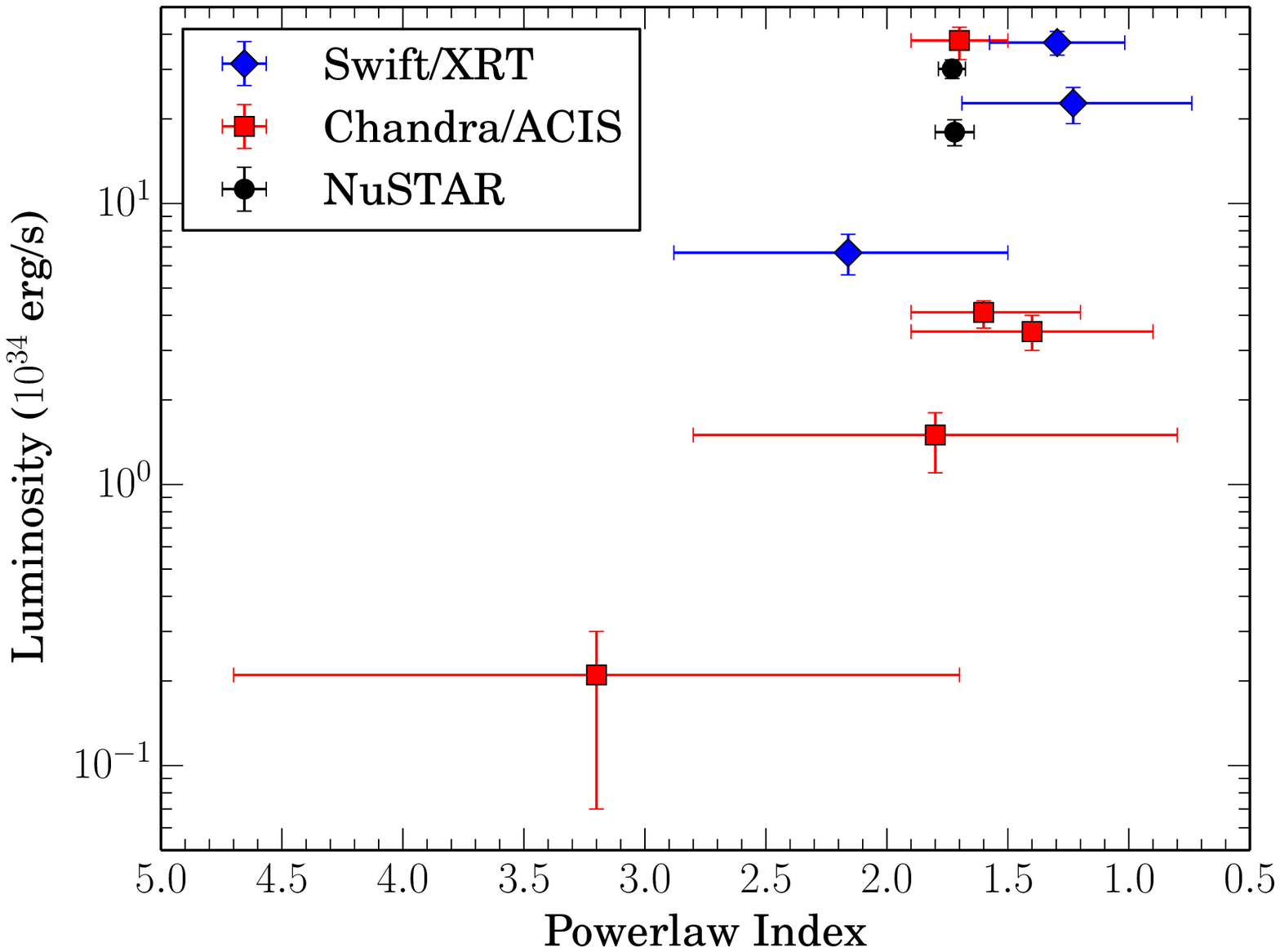}
\caption{\emph{Left:} The light curve of the outburst using \textit{Swift/XRT}(blue diamond), \textit{Chandra/ACIS}(red squares) \& \textit{NuSTAR}(black circles) observations. \textit{Swift/XRT} data points that cover more than one day are merged observations. Note that only statistical uncertainties are included in the X-ray error bars. \emph{Right:} The spectral evolution of the outburst shown with \textit{Chandra/ACIS}(red), \textit{NuSTAR}(black), and \textit{Swift/XRT}(blue).}
\label{fig:lcurve} 
\label{fig:spectral_evo} 
\end{figure*}

The Swift/XRT lightcurve, due to its regular, frequent sampling, is essential for understanding the overall shape of the 2013 outburst of \j174540 (Figure \ref{fig:lcurve}).  The short Swift/XRT observations do not provide precise measurements of the photon index as provided by the other two X-ray instruments, though they are generally consistent.  We combined Swift/XRT spectra from adjacent days (see MJD column of Table 2) to obtain the best possible constraints.  There are hints of variations in the hardness (quantified by the fitted photon index; Table 2, Figure 1 right) in the Swift/XRT data, although these are not confirmed by the longer and more sensitive \Chandra\ and \Nustar\ observations, often taken at, or nearly at, the same time.  The faintest Swift/XRT measurements (data after MJD 56441) are too heavily contaminated by \magnetar\ to provide useful spectra.   See \S \ref{sub:previous_outbursts} below for comparison to previous outbursts of this source.

\subsection{Chandra/ACIS} 
\label{sub:chandra}

The \textit{Chandra/ACIS} combination gives high resolution imaging and spectroscopy, with 0.5'' spatial resolution and typically $<0.3$ keV energy resolution, over the energy range 0.5-10.0 keV \citep{Weisskopf02}.  The High-Energy Transmission Grating (HETG) can be placed into the optical path, diffracting $\sim$50\% of the photons from a point source into spectra dispersed by energy on the detector, with very high energy resolution \footnote{http://cxc.harvard.edu/proposer/POG/html/chap8.html} .  The High Energy Grating (HEG) and Medium Energy Grating (MEG) are optimized for different energy ranges, while the undiffracted photons produce a zeroth-order image on the detector.

We observed the region surrounding Sgr A* regularly from 25 May 2013 to 27 July 2013 for a combined Chandra exposure of 132 ks over 6 observations (see Table \ref{tab:observs}). Three observations (Obs IDs: 15040, 15651, 15654) were performed using \textit{Chandra/ACIS} with a 1/2 sub-array (to reduce the readout time for the CCD) and HETG, due to the brightness of \magnetar\ and \j174540.
This enabled a dispersed grating spectrum of \j174540\ to be extracted for the initial observation on 25 May. However, \j174540\ had dimmed significantly in the subsequent observations, and the dispersed grating spectra were not of sufficient S/N to be usable. The other three observations (Obs ID: 14703; Obs ID: 14946; Obs ID: 15041) were taken with \textit{Chandra/ACIS} in a 1/8 sub-array. Archival Chandra data was also reanalyzed, from the 2003 \citep{Muno05} and 2006 \citep{Degenaar09} outbursts (see Table 1), for an additional 30 ks of observation time. 

All \Chandra\ data were reduced and analyzed using CIAO 4.5 \citep{Fruscione06} following the CIAO science threads\footnote{http://cxc.harvard.edu/ciao/threads}. Each data set was reprocessed using the CIAO routine \textit{chandra\_repro}. Undispersed spectra from each observation were extracted using the CIAO tool \textit{specextract}. 
Source extraction regions were optimized for maximal S/N ratio, and background was extracted from a surrounding annulus when the source was faint, or a nearby region when bright.  

In ObsID 15041 (45 ks long), \j174540\ was not detected.  We estimate a count excess of 7.7$^{+10.2}_{-8.0}$ counts over background (2-5 keV) at this location, which we conservatively interpret as an upper limit of 18 counts (at 90\% confidence).  We calculate an intrinsic $L_X$ limit from our countrate limit using either a $\Gamma$=1.7 power-law, or an NSATMOS hydrogen-atmosphere neutron star model (assuming a 1.4 $M_{\odot}$ and 10 km NS) \citep{Heinke06a}, finding in either case $L_X$(2-10 keV)$<$$4\times10^{32}$ ergs/s. This gives a temperature of $< 1.9\times10^6$ K and indicates that \j174540\ has returned to quiescence.

Dispersed grating spectra were extracted with the CIAO routine \textit{tgextract} as per the HETG/ACIS science thread\footnote{http://cxc.harvard.edu/ciao/threads/spectra\_multi\_acis}, taking care not to include the dispersed spectra from \magnetar\ in the spectra or backgrounds for \j174540. The first order grating spectra (HEG and MEG) were combined using the CIAO routine \textit{add\_grating\_orders}.  The +1 and -1 order response matrix files (RMFs) were found to be indistinguishable, so we used the -1 order RMFs for both. 
We analyzed the first order grating spectra (a combined first-order HEG spectrum, and combined first-order MEG spectrum) from the \textit{Chandra/ACIS} ObsID 15040.  Unfortunately, the total counts were only 860 and 745 in the combined MEG first-order and HEG first-order spectra respectively, with less than 40 counts in each higher-order spectrum (hence, we ignore the higher-order spectra).  Furthermore, the HEG spectrum of \j174540\ suffers from a dispersion angle that places it quite close to that of \magnetar, and overlapping with the complex around Sgr A*, making background subtraction difficult.  

We first bin the grating spectra by a factor of 10 in wavelength, undersampling the spectral resolution by a factor of two, to look for any signs of emission or absorption lines.  We use an absorbed powerlaw model (as fit to the undispersed spectra) for our continuum.  
Although there are weak suggestions of emission lines at 4.5, 4.9, and 6.4 keV in the combined MEG spectrum (none significant), these energies show only suggestions of absorption in the combined HEG spectrum.  We therefore provide only upper limits on the flux from narrow ($\sigma$=0.1 keV) emission lines at these locations, using the MEG spectrum; at 6.4 keV, $F_X<$1.4$\times10^{-12} $ erg  cm$^{-2}$ s$^{-1}$ (Equivalent width of $<0.07$ keV); at 4.5 keV, $F_X<$1.8$\times10^{-12} $ erg  cm$^{-2}$ s$^{-1}$ (Equivalent width of $<0.09$ keV), or at 4.9 keV, $F_X<$1.9$\times10^{-12} $ erg  cm$^{-2}$ s$^{-1}$ (Equivalent width of $<0.06$ keV).
We also group the data to require 20 counts/bin, and perform independent fits to the continuum of the MEG and HEG data.  We find values for the photon index ($1.5\pm0.9$ or $1.9\pm1.0$, respectively) and flux ($5.2^{+1.2}_{-0.8}\times10^{-11}$ or $4.4^{+1.2}_{-0.9}\times10^{-11}$ erg cm$^{-2}$ s$^{-1}$) from the MEG and HEG data that are  consistent with (but with larger errors than) the fit to the simultaneous zeroth-order undispersed spectrum (Table 2).  

All of the \Chandra\ spectra (undispersed, and dispersed) are consistent with an absorbed power-law of photon index consistent with 1.7 (see Table 2). Pileup effects were considered when modeling the bright undispersed spectra, as the peak count rate reaches $\approx$0.2 ct/s, even though the data were taken using subarrays (see \S\ref{sub:outburst}). 
The \Chandra\ undispersed spectra provide the highest-quality data at low energies, which we use below to obtain the best estimate of $N_H$ (\S 3.2).  The \Chandra\ data also provide the highest-precision position for \j174540\ (\S 3.1).

\subsection{VLA}

We carried out radio continuum VLA imaging of the Sgr A* region on May 25, 2013, simultaneous with \Chandra\ ObsID 15040.  A detailed account of these observations will be given elsewhere.  Briefly, radio data were taken with the VLA in its BnC configuration at 0.7 cm.  We used the new correlator setup employing 2 GHz bandwidth around 43 GHz. 
The initial calibration was done with OBIT \citep{Cotton08} employing the phase calibrators J1744-3116,  J1733-1304  and the absolute calibrator 3C286. 
Imaging was performed in AIPS  with a resolution of 1.3$"\times1.1''$ (PA$\sim -89^{\circ}$).  Unfortunately \j174540\ is projected against the edge of the northern arm of the minispiral \citep{Ekers83}, producing a very high radio continuum background, and \j174540\ was not detected in this observation.    
The rms noise of the image per beam, after phase and amplitude self-calibration, is $\sim$2 mJy. 

Different correlations between radio and X-ray luminosities have been observed for black holes \citep{Corbel03,Gallo06,Gallo12} vs. for neutron stars \citep{Migliari06}.  Thus, measuring the radio flux of an unknown X-ray luminous object can help to determine its nature.  However, the radio upper limit we derive here is rather shallow.  For a distance of 8 kpc, we calculate a radio luminosity 1-$\sigma$ upper limit (up to 8.5 GHz, assuming a flat spectrum) of $3.3\times10^{30}$, compared with a simultaneous \Chandra\ X-ray measurement of $L_X$(2-10 keV)=$4\times10^{35}$ ergs/s.  This upper limit is well above the radio detections of similarly-luminous neutron star X-ray binaries \citep{Migliari06}, and also above the radio detections of all similarly luminous black hole X-ray binaries \citep[e.g.][]{Gallo12}.  Thus, due to the high continuum background at this location and the relative faintness of this transient, we cannot draw conclusions from the radio nondetection, and cannot distinguish whether this VFXB contains a neutron star or a black hole.

\begin{table*}
\begin{center}
\caption{\bf X-ray observations of \j174540\ during 2013.}
\begin{tabular}{lllll}
\hline
Observation ID & Date  		& Exposure Time & Luminosity 		& Notes$\star$ \\
            		&   	    		&                          &     ($\times10^{34}$ erg s$^{-1}$)  & \\
\hline
\hline
91736031       & 2013-05-11 & 957 s         & $<1.6$ &     \\
N-2013006 & 2013-05-11 & 32.6 ks	& $<$0.5    &  \\
91736032       & 2013-05-12 & 926 s         & $<2.1$ &     \\
91736033       & 2013-05-13 & 958 s         & $4.0\pm1.6 $ & ~     \\
91736035       & 2013-05-15 & 941 s         & $6.4\pm1.6 $ &  1  \\
91736036       & 2013-05-16 & 987 s         & $6.6\pm1.6 $ & 1   \\
N-2013008 & 2013-05-18 & 38.8 ks	& $35\pm2$	& 2\\
91736037       & 2013-05-18 & 956 s         & $28\pm4$ & 1   \\
91736038       & 2013-05-19 & 952 s         & $21\pm3$  & 1  \\
91736039       & 2013-05-20 & 1006 s        & $25\pm3$  & ~     \\
91736040       & 2013-05-21 & 978 s         & $34\pm3$ & ~     \\
91736041       & 2013-05-22 & 1005 s        & $34\pm3$ & ~     \\
91736042       & 2013-05-23 & 958 s         & $39\pm4$ & ~     \\
C-15040              & 2013-05-25          & 24.4 ks        & 38$^{+4}_{-3}$  & 3  \\
N-2013010 & 2013-05-27 & 37.3 ks	& $18\pm2$	& ~ \\
91712006       & 2013-05-29 & 956 s         & $7\pm2 $ & ~     \\
91736044       & 2013-05-30 & 972 s         & $4.8\pm2 $ & ~     \\
91736045       & 2013-05-31 & 991 s         & $5\pm2 $ & ~     \\
91736046       & 2013-06-01 & 872 s         & $6\pm2 $ & ~     \\
91736047       & 2013-06-02 & 1036 s        & $5\pm2 $ & ~     \\
91736048       & 2013-06-03 & 1044 s        & $5\pm1 $ & ~     \\
91736049       & 2013-06-04 & 1007 s        & $5\pm1 $ & ~     \\
C-14703              & 2013-06-04          & 18.6 ks        & 4.1$^{+0.5}_{-0.4}$ & ~     \\

91712007       & 2013-06-05 & 980 s         & $4\pm2$ & ~     \\
C-15651              & 2013-06-05          & 14.1 ks        & 3.5$\pm0.5$ & ~     \\
91736052       & 2013-06-08 & 1119 s        & $2\pm1 $ & ~     \\
91736053       & 2013-06-09 & 947 s         & $<2.0$ & ~     \\
C-15654              & 2013-06-09          & 9.3 ks         & 1.5$^{+0.4}_{-0.3}$ & ~     \\
91736054       & 2013-06-10 & 957 s         & $<1.2$ &     \\
91736056       & 2013-06-13 & 1250 s        & $2\pm1$ &    \\
91736057       & 2013-06-14 & 941 s         & $2\pm1$ &     \\
N-2013012 & 2013-06-14 & 26.8 ks	& $<$3.2 &  \\
91736061       & 2013-06-18 & 992 s         & $<$2.0 &     \\
91712009       & 2013-06-19 & 1067 s        & $<$2.5 &     \\
91736062       & 2013-06-20 & 1020 s        & $<$3.2 &     \\
91736063       & 2013-06-21 & 1079 s        & $<$4.8 &     \\
91736064       & 2013-06-24 & 1050 s        & $<$3.2 &    \\
91736065       & 2013-06-25 & 1085 s        & $<$3.7&     \\
91712010       & 2013-06-26 & 756 s         & $<$2.6 &    \\
91736066       & 2013-06-28 & 1001 s        & $<$2.6 &     \\
91736068       & 2013-06-29 & 1146 s        & $<$3.4 &     \\
91736069       & 2013-06-30 & 937 s         & $<$4.0 &     \\
91736070       & 2013-07-01 & 955 s         & $<$2.9 &     \\
91736071       & 2013-07-02 & 955 s         & $<$4.2 &     \\
C-14946        & 2013-07-02  & 20.1 ks       & 0.38$^{+0.23}_{-0.15}$ &   \\
91712011       & 2013-07-03 & 1061 s        & $<$4.0 &     \\
91736073       & 2013-07-05 & 1019 s        & $<$4.2 &     \\
91736074       & 2013-07-06 & 970 s         & $<$4.5 &     \\
C-15041         & 2013-07-27 & 45.4 ks     & $<$0.04 &  \\
\hline
\end{tabular}
\end{center}
\smallskip
List of observations used in our analysis, with 2-10 keV $L_X$ estimates. ObsIDs have a C- for Chandra, N- for \Nustar\ (omitting the 8000 at the beginning of each \Nustar\ ObsID), or refer to Swift/XRT observations.  $\star$  References to previous analyses of these observations: - (1) \citet{Degenaar13e}; (2) \citet{Dufour13}; (3) \citet{Heinke13_atel}.
\label{tab:observs}
\end{table*}

\begin{table*}
\caption{\bf Fits to 2003, 2006, \& 2013 X-ray spectra of \j174540}
\begin{center}
\begin{tabular}{llllllll}
Observation ID & MJD  & $N_H$ & Photon Index & Flux & $L_X$ (2-10 keV) & D.O.F. & Red. $\chi^{2}$ \\
 & & ($\times10^{22} $cm$^{-2}$) & & ($\times10^{-12} $erg  cm$^{-2}$ s$^{-1}$) & (erg s$^{-1}$) \\
\hline
\hline
 \multicolumn{8}{c}{2003} \\
\hline
C-03549   &   52809.8       & 15.1 (fixed) & 1.6$\pm0.2$   & 6.6$\pm0.3$ &   5.1$\pm0.2\times10^{34}$   &  26       &  0.93 \\ 
\hline
\hline
 \multicolumn{8}{c}{2006} \\
\hline
3564978-84       & 54023.0-54029.0 & 15.1 (fixed)   &   1.9$\pm0.6$   &   4.1$\pm0.7$   & 3.3$\pm0.5\times10^{34}$  &   32   &   0.88   \\
3564985          & 54030.1 &    t   &   1.5$^{+0.6}_{-0.7}$   &   19.4$^{+4.0}_{-3.6}$ &   1.5$\pm0.3\times10^{35}$   &   t   &   t   \\
3564986-87       & 54031.1-54032.0 &    t   &   1.7$\pm0.5$   &   13.4$\pm2.0$   &  1.0$\pm0.2\times10^{35}$   &   t   &   t   \\
3564988-92       & 54032.9-54039.5 &    t   &   2.0$^\pm0.4$   &   5.6$\pm0.6$ &  4.3$\pm0.5\times10^{34}$  &   t   &   t   \\ 
06646   &   54037.1       & t & 1.7$\pm0.6$ & 2.4$\pm0.4$ &  1.8$\pm0.3\times10^{34}$  &  t     &  t   \\ 
\hline
\hline
\multicolumn{8}{c}{2013} \\
\hline
C-15040   	&  56437.5  & 15.1$^{+1.6}_{-1.4}$        &       1.7$\pm0.2$        &      49.7$^{+5.5}_{-4.4}$ &     3.8$^{+0.4}_{-0.3}\times10^{35}$     &      178         &   0.75        \\
C-14703   					&  56447.4  &   t        &      1.6$^{+0.4}_{-0.3}$         &    5.4$^{+0.6}_{-0.5}$ &     4.1$^{+0.5}_{-0.4}\times10^{34}$     &       t        &    t       \\
C-15651    					&  56448.9  &   t        &      1.4$\pm0.5$        &      4.6$\pm0.6$ &     3.5$\pm0.5\times10^{34}$     &       t        &    t       \\
C-15654    					&  56452.2  &   t        &      1.8$\pm1.0$         &     2.0$^{+0.5}_{-0.4}$ &    1.5$^{+0.4}_{-0.3}\times10^{34}$      &       t        &    t       \\
C-14946    					&  56475.3  &   t        &      3.2$^{+1.5}_{-1.5}$         &    0.29$^{+0.18}_{-0.13}$  &  2.1$^{+1.4}_{-0.9}\times10^{33}$  &       t        &    t       \\
91736031-36             & 56423.4-56428.1  &  15.1 (fixed)   &   2.2$^{+0.8}_{-0.9}$  	&  6.1$\pm1.3$ 		&  4.6$\pm0.8\times10^{34}$   		&   43       &   0.64 \\
91736037-39             & 56430.8-56432.0  &  t      		&   1.2$^{+0.4}_{-0.5}$      &  31$\pm4$  			&  2.4$\pm0.3\times10^{35}$    		&  t       &  t       \\
91736040-42             & 56433.1-56435.3  &  t    		&   1.2$\pm0.3$       		&  50$\pm5$ 			&  3.8$\pm0.4\times10^{35}$    		&  t       &  t       \\
91736044-47 \& 91712006 & 56441.5-56445.7  &  t      	&   2.0$\pm0.7$       		&  8$\pm1$ 			&  6$\pm1\times10^{34}$   		&  t       &  t       \\
91736048-57 \& 91712007 & 56446.6-56457.2  &  t      	&   2.1$\pm0.7$       		&  4.6$^{+0.8}_{-0.7}$ 	&  3.4$^{+0.7}_{-0.6}\times10^{34}$  &  t       &  t       \\
N-2013008 		&  56430.7  & 25$\pm$3	&	1.73$\pm$0.08	&	39$\pm$3		& $3.5\pm0.2 \times 10^{35}$	& 467  & 1.01 \\
N-2013010 		&  56439.4  & 27$\pm$5	&	1.72$\pm$0.08	&	23$_{-2}^{+3}$	& $1.8\pm0.2 \times 10^{35}$	& 337  & 0.99 \\

\hline
\hline
\end{tabular}
\end{center}
\smallskip
Joint and individual fits of the \textit{Swift/XRT}, \textit{Chandra/ACIS} (indicated with C-) and \textit{NuSTAR} (indicated with N-) data sets. Luminosities are calculated using a distance of 8.0 kpc. The luminosity and flux are calculated over the 2-10 keV range. 
\label{tab:group_fits}
\end{table*}

\begin{table*}
\caption{\bf Comptonization Model Fits to NuSTAR Spectra of \j174540}
\begin{center}
\begin{tabular}{lccccc}
\hline
Obs. ID			&	N$_H$			&	T$_0$			&	kT			&	$\tau$               	   				&  $\chi^2$/d.o.f \\
     				&(10$^{22}$ cm$^{-2}$)	&	(keV)			&  (keV)			&				      	   &	      \\
\hline
2013008	&	16$_{-5}^{+6}$		&	1.1$_{-0.3}^{+0.2}$ & $>$40 &  0.05$_{-0.01}^{+1.28}$ & 	469.72/465 \\
2013010	&	21$_{-9}^{+8}$		&  	0.9$_{-0.9}^{+0.3}$	& 14$_{-3}^{+96}$	 &  3.04$_{-2.96}^{+0.7}$   & 331.84/335 \\
\hline
\end{tabular}
\end{center}
\smallskip
Details of {\it TBABS*COMPTT} spectral modeling of the two NuSTAR observations of \j174540.
\label{tab:comptt}
\end{table*}

\section{Key Results} 
\label{sec:results}

\subsection{Position} 
\label{sub:position}

The position of the new transient was derived using \textit{Chandra/ACIS} ObsID 15040, in which both \magnetar\ and the transient were clearly detected using the {\it wavdetect} program \citep[Figure \ref{fig:images}; see also][]{Heinke13_atel}. The position of \magnetar\ was calculated by \citet{Shannon13}, using an ATCA radio interferometric observation, to be RA= 17:45:40.16$\pm$0.022" and Dec= -29:00:29.82$\pm$0.09". We thus shifted our {\it Chandra} positions to find the astrometrically corrected position of the new transient, RA= 17:45:40.07$\pm$0.1" and Dec= -20:00:05.8$\pm$0.1".  This result agrees with the published position of \j174540,  RA= 17:45:40.06$\pm$0.6" and Dec= -29:00:05.5$\pm$0.6" \citep{Muno05,Degenaar09}, confirming the identification. We adopt this updated, boresited position for \j174540\ for our entire analysis. The close proximity of \j174540\ to Sgr A*, 23" or 0.9 parsecs in projection, given the  concentration of X-ray sources  and transients close to Sgr A* \citep{Muno03,Muno05}, strongly indicates that the distance to \j174540\ is essentially identical to that of Sgr A*, for which we adopt 8 kpc \citet{Reid93}.  The high absorption, consistent with other sources near Sgr A* \citet{Muno03}, agrees with the adopted distance.

\subsection{Column Density} 
\label{sub:outburst}

Since the individual spectra were typically not sufficient to effectively constrain the column density, we first conducted joint spectral fits designed to measure $N_H$. Two joint fits were performed (Table \ref{tab:group_fits}), one for the 6 groups of \textit{Swift/XRT} data, and the other for the \textit{Chandra/ACIS} data, due to the differing fractions of the dust scattering halo that would be encompassed in the spectral extractions from the two instruments. Dust scatters a fraction of the X-rays out of the line of sight, typically at small angles \citep[e.g.][]{Predehl95}.  For the expected distribution of dust towards Galactic Center sources (much of the dust within 100 pc), $\sim$50\% of the scattered 2 keV X-rays should be found within a 1'' scattering halo of Sgr A* \citep{Tan04}, whereas virtually all the scattered halo should lie within the Swift/XRT extraction region.  We used the photoionization cross sections given by \citet{Verner96}, and abundances from \citet{Wilms00}, in our photoelectric absorption model.

We used an absorbed powerlaw to model the data, taking into account scattering with the SCATTER model \citep{Predehl03}, (PILEUP $\times$ TBABS $\times$ PEGPWRLW $\times$ SCATTER), to model the \textit{Chandra/ACIS} spectra.  The dust extinction parameter A$_V$ was set to n$_H$/0.177 and $\alpha$ in the pileup model was frozen to 0.5 as per \citet{Davis01}.  Freeing the pileup parameter $\alpha$ did not substantially alter our results.

Since our extraction region was several times the expected scattering halo size for the \textit{Swift/XRT} and \textit{NuSTAR} spectra, and pileup was not a concern, we simplified our model for these data to TBABS$\times$PEGPWRLW. The power-law photon indices and flux normalizations for each observation were left untied during fitting.
The column density value from our joint \textit{Chandra/ACIS} fit, N$_H$=15.1$^{+1.6}_{-1.4}\times10^{22}$cm$^{-2}$, was well-constrained. The column density value for the \textit{Swift/XRT} fit was N$_H$=12.6$^{+3.5}_{-3.0}\times10^{22}$cm$^{-2}$, consistent with the \textit{Chandra/ACIS} value but with larger errors.  The neutral $N_H$ column in low-mass X-ray binaries of low to moderate inclination seems empirically to remain constant during outbursts \citep{Miller09}, except for the highest $L_X$ outbursts \citep{Oosterbroek97, Zycki99}.
We therefore used the \textit{Chandra/ACIS} measurement as our $N_H$ value in analyses of individual Chandra and Swift spectra.  The \textit{NuSTAR} spectral fits require  a larger value of $N_H$=$2.6\times10^{23}$ cm$^{-2}$ for power-law fits.  We note that although the formal uncertainties are small ($3\times10^{22}$ cm$^{-2}$ for the first ObsID), there are complexities in the NuSTAR background subtraction at low energies (where \magnetar\ dominates the flux). 
The $N_H$ values obtained from these fits are consistent with a location at the Galactic Center, as expected given the close proximity to Sgr A*.

\subsection{Lightcurve and Spectral Evolution}

Figure \ref{fig:spectral_evo} (left) shows the lightcurve of \j174540\ over the entire outburst.  It rose from $L_X$(2-10 keV)$<$$5\times 10^{33}$ ergs/s on May 11 up to $4.0\times10^{34}$ ergs/s on May 13, and then to a peak of $3.8\times10^{35}$ ergs/s on May 18.  The lightcurve shows evidence for a dip of almost a factor of two for a few days, then recovers and reaches a briefly higher peak on May 25, at $5.0\times10^{35}$ ergs/s, before rapidly declining by a factor of seven over four days.  The lightcurve then decays more slowly, by a factor of two over ten days, before falling below Swift/XRT's detection limit. 

 A final \Chandra\ observation another 24 days later finds \j174540\ another factor of $\sim$7 fainter, at 2.1$^{+1.4}_{-0.9}\times10^{33}$ erg s$^{-1}$, and possibly with a softer spectrum (the best-fit photon index was 3.2$^{+1.5}_{-1.5}$, not significantly different than at higher $L_X$). 
The Swift/XRT upper flux limit, averaged over the last 20 days, is slightly lower than the flux from the last Chandra observation (near the end of those 20 days), suggesting that the flux measured in the final Chandra observation was not persistent throughout that time period.
It is clear that the outburst as a whole did not last longer than 26 days above $4\times10^{34}$ ergs/s, with a total 2-10 keV (unabsorbed) fluence of $\sim4.7\times10^{-5}$ ergs cm$^{-2}$, or for an 8.0 kpc distance (considering its proximity to Sgr A*), a total 2-10 keV emitted energy of $\sim3.6\times10^{41}$ ergs.

An absorbed powerlaw model adequately fits all the datasets (see Table 2, and Figure \ref{fig:xspec}).  
We plot the variation in the fitted power-law photon index (chosen as a more instrument-independent quantity than hardness ratios) in Figure \ref{fig:spectral_evo} (right).  The data are consistent with a fixed photon index of $\Gamma=1.7$ for all observations. 
 There is a slight suggestion of spectral evolution in the Swift/XRT data alone (Figure \ref{fig:spectral_evo}, Table \ref{tab:group_fits}), indicating softening during the decline from $2\times10^{35}$ ergs/s and $6\times10^{34}$ ergs/s.  However, this is not supported by the (higher-statistics) Chandra data, which show no evidence of spectral changes between observations at the peak $L_X$ ($4\times10^{35}$ ergs/s) down to $1.5\times10^{34}$ ergs/s.  The average fitted photon indices from the Chandra, Swift/XRT, and NuSTAR data agree quite well.

\subsection{Comparison to Previous Outbursts} 
\label{sub:previous_outbursts}

The Galactic Center has been monitored regularly using the Swift/XRT telescope since 2006, sometimes on a daily basis, in other years typically every third day \citep{Degenaar09,Degenaar13c}.  Two outbursts of \j174540\ have now been monitored by Swift/XRT \citep{Degenaar09}, while a glimpse of a previous outburst was provided serendipitously by \Chandra\ \citep{Muno05}.  The 2006 outburst lasted for roughly two weeks with a peak 2-10 keV $L_X$ of $2\times10^{35}$ ergs/s \citep{Degenaar09}, thus giving a smaller fluence of $1.3\times10^{-5}$, vs. our measured fluence of $4.7\times10^{-5}$ ergs cm$^{-2}$. 
The single 2003 \Chandra\ observation found \j174540\ at 3.4$\times10^{34}$ erg s$^{-1}$ (2-8 keV) \citep{Muno05}.  If this outburst resembled the two better-studied outbursts of this source, the 2003 observation likely occurred during the outburst tail.
We have analyzed the spectra of the previous Chandra and Swift/XRT observations, and find that the spectral parameters (photon index specifically) are consistent with those of our observations (Table 2).

We can re-assess the duty cycle and average mass transfer rate of \j174540\ using the full Swift/XRT campaigns.  The total Swift/XRT 2006-2013 campaigns covered roughly 270 weeks.  Since the two outbursts observed by Swift/XRT each lasted over 2 weeks, we believe that a similar outburst would have been caught by these observations, which occurred every 1-4 days (excepting periods of $\sim$2 months each year when Sgr A* is too close to the Sun for observations).  Thus, we estimate a duty cycle of (2+3)/270$\sim$2\%, consistent with the 1-6\% range estimated by \citet{Degenaar09}.  Similarly, the average mass transfer rate, assuming a bolometric luminosity 3 times the 2-10 keV $L_X$ \citep{intZand07}, and a 1.4 \Msun, 10 km NS, comes to $\sim7\times10^{-14}$ \Msun/year, somewhat smaller than the range inferred by \citet{Degenaar09}.

\section{Discussion} 
\label{sec:discussion}

\subsection{Spectra of \j174540}

Few LMXBs with $L_X$ (0.5-10 keV) between $10^{34}$ and $3\times10^{35}$ ergs/s have high-quality spectra above 10-20 keV, such as our \Nustar\ spectra provide.  
Such spectra may enable us to probe how the ``hard'' accretion state in these systems, which can typically be fit by power-law models without a cutoff up to nearly 100 keV, transitions to the quiescent state.  \citet{Chakrabarty14} have recently shown that the nonthermal component in Cen X-4 shows significant curvature in quiescence, being well-fit by a bremsstrahlung spectrum of kT=18$\pm1$ keV.
Possibly the highest-quality data on the hard X-rays during the transition to quiescence are the RXTE (PCA and HEXTE) spectra of SAX J1808.4-3658 (SAX J1808) in its 1998 and  2002 outbursts, since SAX J1808 is not crowded and is relatively nearby (3.5 kpc).  
\citet{Gilfanov98} and \citet{Gierlinski02} presented 1998 RXTE spectra of SAX J1808, evolving from $L_X=3\times10^{36}$ ergs/s down to $3\times10^{35}$ ergs/s between 3 and 100 keV, and then dropping down to $3\times10^{34}$ ergs/s (only to 20 keV, \citealt{Gilfanov98}), while \citet{Ibragimov09} presented 2002 RXTE spectra declining from $3\times10^{36}$  down to $3\times10^{35}$ ergs/s between 3-100 keV.  These authors all find that SAX J1808's spectrum remained essentially constant as it declined by over an order of magnitude down to $L_X\sim3\times10^{35}$ ergs/s, though they were unable to clearly detect it above 20 keV below this luminosity.  
\citet{Gierlinski02} use a Comptonization model for their fits, and find $kT_e=90^{+240}_{-30}$ keV for a time-averaged spectrum, although they are unable to measure how $kT_e$ may vary across their $L_X$ range.  Similarly, \citet{Ibragimov09} find that a Comptonization model for the higher-$L_X$ spectra requires  $kT_e \sim$40-45 keV, while it is unconstrained for $L_X=3\times10^{35}$ ergs/s.  
The LMXB with the faintest peak luminosity to have spectral information above 10 keV is perhaps IGR J17285-2922, a VFXB discovered towards the Galactic bulge with INTEGRAL \citep{Walter04}.  Its two known outbursts have reached (assuming an 8 kpc distance) $L_X$(2-10 keV)$\sim$$5\times10^{35}$ ergs/s (in 2003, \citealt{Barlow05}), and $L_X$(2-10 keV)$\sim$$4\times10^{35}$ ergs/s \citep[in 2010,][]{Sidoli11}.  Both outbursts were observed by INTEGRAL (the second also by XMM), and showed hard time-averaged spectra ($\Gamma$ of $2.1\pm0.2$ and $1.60\pm0.01$ respectively) extending out beyond 100 keV, without evidence of a spectral turnover (though these INTEGRAL/ISGRI  spectra are not high S/N, due to the source faintness).

Thus, our lower-luminosity \Nustar\ spectrum (ObsID 2013010, see \S 2.1, Figure \ref{fig:xspec}) is the lowest-luminosity spectral measurement of an LMXB transient in outburst (that we are aware of) above 20 keV.  
We do not see clear evidence for a spectral turnover below 80 keV, nor do we see differences with higher-$L_X$ spectra.  This observation clearly shows that \Nustar\ can provide excellent hard X-ray spectra of NS X-ray binary transients at even fainter luminosities, if they are closer to Earth (e.g. SAX J1808), allowing us to see how the hard spectra evolve at lower luminosities.  

\section*{Acknowledgments}
We are very grateful for the hard work by the staff and directors of the observatories, and in particular, to Scott Wolk, Patrick Slane, and Dillon Foight at the CXC, for enabling our monitoring campaigns of the Galactic Center. 
COH is supported by an NSERC Discovery Grant and an Alberta Ingenuity New Faculty Award.  
DH is supported by {\it Chandra} X-ray Observatory (CXO) Award Numbers GO3-14121X, GO3-14099X, and G03-14060X, operated by the Smithsonian Astrophysical Observatory for and on behalf of NASA under contract NAS8-03060, and also by NASA {\it Swift} grant NNX14AC30G.  
NR is supported by a Ramon y Cajal Fellowship, NWO Vidi award, and grants AYA 2012-39303, SGR2009-811, and iLINK 2011-0303.
ND is supported by NASA through Hubble Postdoctoral Fellowship grant number HST-HF-51287.01-A from STScI.  
This research made use of data obtained from the {\it Chandra} Data Archive and software provided by the {\it Chandra} X-ray Center in the application packages CIAO, ChIPS, and Sherpa.  The National Radio Astronomy Observatory is operated by Associated Universities Inc., under cooperative agreement with the National Science Foundation.
The \Swiftxrt\ Data Analysis Software (XRTDAS) is developed under the responsibility of the ASI Science Data Center (ASDC), Italy.\\
We acknowledge extensive use of the ADS and arXiv.

\bibliographystyle{mn2e}
\bibliography{ref_list_general}

\bsp

\label{lastpage}

\end{document}